\documentclass[prl,aps,twocolumn,showpacs]{revtex4}
\usepackage{epsfig}
\usepackage{graphicx}
\usepackage{textcomp}

\begin{document}
\title{Noise driven unlimited population growth}

\author{Baruch Meerson$^{1}$ and Pavel V. Sasorov$^{2}$}

\affiliation{$^{1}$Racah Institute of Physics, Hebrew University
of Jerusalem, Jerusalem 91904, Israel}

\affiliation{$^{2}$Institute of Theoretical and Experimental
Physics, Moscow 117218, Russia}


\begin{abstract}
Demographic noise causes unlimited population growth in a broad class of models which, without noise, would predict a stable finite population.  We study this effect on the example of a stochastic birth-death model which includes immigration, binary reproduction and death. The unlimited population growth proceeds as an exponentially slow decay of a metastable probability distribution (MPD) of the population. We develop a systematic WKB theory, complemented by the van Kampen system size expansion, for the MPD and for the decay time. Important signatures of the MPD is a power-law tail (such that all the distribution moments, except the zeroth one, diverge) and the presence in the solution of two different WKB modes.

\end{abstract}

\maketitle

Since the celebrated essay of Malthus \cite{Malthus} quantitative modeling of population dynamics has attracted much interest. To a large extent, this interest is powered by the danger of a Malthusian catastrophe, when too a rapid population growth causes a fatal lack of resources.
Here we focus on a variant of Malthusian catastrophe by considering not too a large population that undergoes binary reproduction, immigration and death. Although macroscopically stable, this population can be pushed to the Malthusian limit (a critical population size that sparks a Malthusian catastrophe) by rare large fluctuations. We show that unlimited population growth proceeds as a slow decay of a metastable probability distribution (MPD) of the population. We determine the MPD and the decay time analytically by developing a systematic WKB theory for the master equation and combining it with the van Kampen system size expansion. We show that the MPD is described by two different WKB modes which are strongly coupled in a narrow region around the  unstable fixed point of the deterministic rate equation of the model. At large population sizes the MPD exhibits a power-law tail so that all the distribution moments, except the zeroth one, diverge.

\textit{Deterministic rate equation}. At the deterministic level of modeling a Malthusian catastrophe does \textit{not} occur if the gain and loss processes balance each other so that the resulting steady-state population size is stable with respect to small perturbations.  Real populations, however, behave stochastically, rather then deterministically \cite{Bartlett}. The stochasticity may cause an unlimited population growth in a broad class of models which deterministic counterparts predict a stable finite population size. We will investigate this remarkable phenomenon on the example of a birth-death model  \cite{Kamenev1} which accounts for binary reproduction $2A\stackrel{\lambda}{\rightarrow} 3A$,
immigration $\emptyset\stackrel{\sigma}{\rightarrow} A$, and death $A\stackrel{\mu}{\rightarrow} \emptyset$. The rate equation for this model is
\begin{equation}\label{rateeq}
   \dot{\bar{n}} = \sigma -\mu \bar{n}+(\lambda/2)\,\bar{n}^2\,,
\end{equation}
where $\bar{n}(t) \gg 1$ is the average population size. For a relatively low death rate, $\mu^2<2\sigma \lambda$, Eq.~(\ref{rateeq}) does not have fixed points, and the population size blows up in finite time for any $\bar{n}(t=0)$. For $\mu^2>2\sigma \lambda$ Eq.~(\ref{rateeq}) has two fixed points: $n_1=\Omega (1 - \delta)$ and $n_2=\Omega (1 + \delta)$, where $\Omega=\mu/\lambda \gg 1$ and $\delta^2=1-2\sigma \lambda/\mu^2$.  When starting from any $\bar{n}(t=0) < n_{2}$, the population size flows to the attracting fixed point $\bar{n}=n_{1}$ with a characteristic relaxation time $\tau_r=1/(\mu \delta)$, and stays there forever.

\textit{Master equation, absorbing state and decay of metastable state.} The demographic noise, ignored by the rate equation (\ref{rateeq}), is accounted for by the master equation, see \textit{e.g.} Ref. \cite{vanKampen}, which governs the evolution of probability $P_n(t)$ to have $n$ individuals at time $t$:
\begin{equation}
\dot{P}_n=\lambda_{n-1} P_{n-1}- (\lambda_n+\mu_n) P_n+\mu_{n+1} P_{n+1}\,,
\label{p10}
\end{equation}
where $\lambda_n=(\lambda/2) \,n (n-1)+\sigma$, and $\mu_n=\mu n$. One striking property of Eq.~(\ref{p10}) concerns its steady state. Summing up the first $n$ equations in Eq.~(\ref{p10}), we obtain
\begin{equation}
\frac{d}{dt}\sum\limits_{l=0}^n P_l(t)=
-\frac{n(n-1)}{2\Omega}P_{n}+(n+1)P_{n+1}-\frac{\gamma \Omega}{2} P_n\,,
\label{p20}
\end{equation}
where $\gamma= 1-\delta^2 = 2\sigma \lambda/\mu^2$, and the time is rescaled by the death rate: $\mu t \to t$. Putting $d/dt=0$, we obtain
\begin{equation}
P_{n+1}= \frac{n(n-1)+\gamma \Omega^2}{2\Omega (n+1)}P_{n}\,.
\label{p50}
\end{equation}
Clearly, $\lim\limits_{n\to\infty} P_n = \infty$ unless $P_n=0$, $n=0,1,2,\dots$. Therefore, the ultimate state of the stochastic process corresponds to an empty system, in a stark contrast to the prediction of the rate equation (\ref{rateeq}). At the same time, there is a \textit{separate} absorbing state at infinity which ``collects" the individuals and ultimately becomes fully populated, as its probability ${\cal P}_{\infty}(t\to \infty) = 1$. Here is an overview of how the unlimited population growth occurs.  At  $t \gtrsim \tau_r$, the MPD sets in, peaked at the attracting fixed point $n=n_{1}$. We will assume (and check \textit{a posteriori}) that the width of the MPD here is much less than the distance $n_{2}-n_{1}$ between the two fixed points of the rate equation. In this regime a \textit{large} fluctuation is needed to bring the population beyond the unstable fixed point $n=n_{2}$ of the rate equation, from where it rapidly escapes to infinity.   As large fluctuations occur with certainty, a full transfer of the population to infinity is certain. At $t \gg \tau_r$ (in the physical units), $P_n(t)$ decays as  $P_n(t) \simeq C \pi_n \exp(-t/\tau)$, whereas ${\cal P}_{\infty}(t)$ grows as ${\cal P}_{\infty}(t) \simeq 1-C \exp(-t/\tau)$. Here $\pi_n$ is the quasi-stationary probability distribution (QSD; it is normalized to unity), and  $C$ is a constant depending on the initial condition: for ``macroscopic" initial conditions $C\simeq 1$. As in other instances of weak-noise-driven escape from a metastable state \cite{vanKampen}, the decay time $\tau$ turns out to be exponentially long compared to the relaxation time $\tau_r$.

The QSD $\pi_n$ and the decay time $\tau$ are determined by the eigenvalue problem
\begin{eqnarray}
  -E \pi_n &=& \frac{1}{2\Omega}\, \left[(n-1)(n-2)\pi_{n-1}-n(n-1)\pi_{n}\right] \nonumber \\
&+&\left[(n+1)\pi_{n+1}-n\pi_n\right]+
\frac{\gamma \Omega}{2}\left(\pi_{n-1}-\pi_n\right),
\label{n20}
\end{eqnarray}
where $E=(\mu \tau)^{-1}>0$ is the rescaled eigenvalue, and we are interested in the eigenmode with the smallest $E$.  We will exploit the large parameter $\Omega \gg 1$ and solve the eigenvalue problem analytically by combining a systematic WKB expansion \cite{Bender} with the van Kampen system size expansion \cite{vanKampen}.

\begin{figure}[ht]
\includegraphics[width=2.0in,clip=]{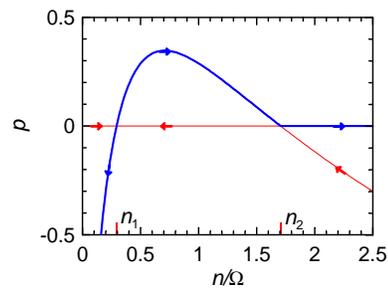}
\caption{(color online). The zero-energy trajectories (\ref{zeroE}) on the phase plane $(n,p)$ for $\gamma=1/2$.
The thick line indicates the WKB modes contributing to the QSD.} \label{fig9}
\end{figure}

\textit{The WKB analysis} of the master equation (\ref{n20}), that we develop here, extends the existing approaches  \cite{Bender,dykman1,Kamenev1,Kessler}, as it accounts for two different WKB modes, and for mode-coupling effects, see below. The WKB ansatz is
\begin{equation}
\pi_n=a(n)\,e^{-S(n)}\,,
\label{n30}
\end{equation}
where, for $n\gg 1$, we can treat the action $S(n)$ and amplitude $a(n)$ as continuous functions of $n$. As can be checked \textit{a posteriori}, the ordering of terms is the following: $S(n)={\cal O}(\Omega)$,  $a(n)={\cal O}(1)$, $S^{\prime}(n)={\cal O}(1)$, $a^{\prime}(n)={\cal O}(1/\Omega)$, $S^{\prime\prime}(n)={\cal O}(1/\Omega)$, $a^{\prime\prime}(n)={\cal O}(1/\Omega^2)$, \textit{etc}. Here and in the following the primes stand for $n$-derivatives. Therefore,  we can approximate
\begin{equation}
\pi_{n\pm1} \simeq \pi_n e^{\mp S^{\prime}}\left(1-\frac{S^{\prime\prime}}{2} \pm\frac{a^{\prime}}{a}\right)\,.
\label{n60}
\end{equation}
Now we substitute Eqs.~(\ref{n30}) and (\ref{n60}) in  Eq.~(\ref{n20}). As $E$ turns out to be \textit{exponentially} small in $1/\Omega$, we must put $E=0$ in all WKB orders.  In the leading order we obtain the eikonal equation $H(n,p)=0$ which describes the trajectories of the time-independent Hamiltonian
\begin{equation}\label{H1}
    H(n,p)=(e^p-1) \left(\frac{n^2}{2 \Omega} -n e^{-p} +\frac{\gamma \Omega}{2} \right)\,.
\end{equation}
Here $n$ is the coordinate, and $p\equiv S^{\prime}$ is the momentum \cite{Kamenev3}. The zero-energy lines,
 \begin{equation}\label{zeroE}
 p=p_s=0\;\;\;\;\mbox{and}\;\;\;\; p=p_f=-\ln \left(\frac{n}{2\Omega}+
\frac{\gamma\Omega}{2n}\right)\,,
 \end{equation}
describe the slow and the fast (as functions of $n$) WKB modes, respectively.  For the slow mode the action $S=0$. One can check that the corresponding Hamilton equation for $n$ coincides with the rate equation (\ref{rateeq}). The fast mode corresponds to the instanton: a heteroclinic orbit of the Hamiltonian (\ref{H1}) which exits the saddle point $(n_1,0)$ and enters the saddle point  $(n_2,0)$ of the phase plane $(n,p)$, see Fig.~\ref{fig9}. In analogy with other problems of noise driven escape \cite{dykman1,freidlin}, this instanton describes the most probable escape path: in this case \textit{to infinity}. The action $S(n)=\int^{n} p_f(n^{\prime}) dn^{\prime}$ along the instanton is
\begin{equation}
S(n) = n  -2\Omega \sqrt{\gamma}\arctan\frac{n}{\Omega\sqrt{\gamma}}
-n \ln \left(\frac{n}{2\Omega}+\frac{\gamma\Omega}{2n}\right)\,,
\label{n170}
\end{equation}
where the integration constant can be put to zero. Note that the saddle points $(n_1,0)$  and $(n_2,0)$
of the Hamiltonian (\ref{H1}) are \textit{mode-crossing} points, as $p_f=p_s=0$ there.

In the subleading order of the WKB expansion
we obtain a first-order equation for the amplitude $a(n)$:
\begin{eqnarray}
&\left[a^2\, \left(n^2e^p-2n\Omega e^{-p}+\gamma\Omega^2e^p\right)\right]^{\prime}&
\nonumber\\
&= 2\left(\Omega e^{-p}-2n e^p+n\right)a^2\,,&
\label{n100}
\end{eqnarray}
For the fast mode we find, after some algebra,
\begin{equation}
a_f(n)= \frac{A_f}{\sqrt{n} \, (n^2+\gamma \Omega^2)}\,,\;\;\mbox{where}\;\; A_f=const\,.
\label{n190}
\end{equation}
Therefore, the fast WKB mode is well behaved. For the slow mode Eq.~(\ref{n100}) yields
\begin{equation}
a_s(n)=\frac{A_s}{(n-n_1)(n-n_2)}\,,\;\;\;\mbox{where}\;\;\; A_s=const\,.
\label{n150}
\end{equation}
The slow-mode solution diverges at each of the two mode-crossing points implying breakdown of the WKB approximation there. To understand the mechanism of breakdown, we notice that it occurs in the regions of small $p\equiv S^{\prime}$, that is a slow variation of $S(n)$ and, therefore, of $\pi_n$. Here we can use the (stationary) Fokker-Planck equation which follows from the van Kampen system size expansion \cite{vanKampen} applied to Eq.~(\ref{n20}) with $E=0$ \cite{FP}:
\begin{equation}
\left[(n-n_1) (n_2-n)\, \pi_n \right]^{\prime}
+
\left\{\frac{1}{2}\left[(n+\Omega)^2-\delta^2\Omega^2\right] \pi_n\right\}^{\prime\prime}=0\,.
\label{n210}
\end{equation}
The first and second terms describes drift and diffusion, respectively.  The mechanism of breakdown of WKB becomes clear once we observe that the slow mode solution $\pi_n=a_s(n)$, as described by Eq.~(\ref{n150}), solves Eq.~(\ref{n210}) with the diffusion term neglected. As we will see shortly, an account of the small diffusion term regularizes the singularity. This regularization is needed only in a narrow boundary layer around the mode-crossing point $n=n_2$. Indeed, at $1\ll n \lesssim n_2$ the slow-mode solution is merely an exponentially small correction to the fast-mode solution, and so it should be discarded there. The situation is different at $n \gtrsim n_2$. Here the fast-mode solution $a_f(n) e^{-S(n)}$ should be discarded, as it diverges as $n \to \infty$, whereas the slow mode yields the correct solution. The slow and fast modes are strongly coupled in the boundary layer around $n=n_2$, and this coupling is described by the boundary-layer solution.

\textit{The boundary layer and asymptotic matching.} Let us consider the stationary Fokker-Planck equation (\ref{n210}) in the vicinity of the mode-crossing point $n=n_2$:  $|n-n_{2}|\ll n_{2}$. Here we can put $n-n_1 \simeq n_{2}-n_{1}$ in the drift term,
and $n\simeq n_2$ in the diffusion term. Integrating the equation once, we obtain
\begin{equation}\label{pi+}
    d \pi(x)/dx - 2 x \,\pi(x) = - C_1\,,
\end{equation}
where $x=(n-n_2)/l_2,\;$ $l_2=\left[2 \Omega (1/\delta+1)\right]^{1/2}$ is the characteristic width of the boundary layer, and $C_1>0$ is a constant. The general solution of Eq.~(\ref{pi+}) is $\pi (x)=C_1 \phi(x) + C_2\, e^{x^2}$,
where $\phi(x) = e^{x^2}\int_x^\infty e^{-\xi^2}\, d\xi$,
and $C_2$ is another constant which, as we will see shortly, must be put to zero. The function $\phi(x)$ has the following asymptotes  at $|x|\gg 1$:
\begin{equation}
\label{n250}
\phi(x)=\left\{\begin{array}{ll}
(2 x)^{-1}+{\cal O}\left(x^{-3}\right)\,, & x\gg 1, \\
 \sqrt{\pi}e^{x^2}+{\cal O}\left(|x|^{-1}\right)\,, & x<0,\,-x\gg 1\,.
\end{array}
\right.
\end{equation}
We start the matching procedure from the region of $n>n_{2}$, where the solution can only include the slow mode (\ref{n150}) with a yet unknown normalization constant $A_s$.  Consider the region $0<n-n_{2} \ll n_{2}$.  In the leading order, Eq.~(\ref{n150}) yields $\pi_n=a_s(n)\simeq A_s/[2\Omega\delta (n-n_{2})]$.
Matching this asymptote with the boundary layer solution $\pi (x)$ in their joint region of validity  $l_{2}\ll n-n_{2} \ll n_{2}$, we obtain $C_1=A_s(\Omega l_2 \delta )^{-1}$ and $C_2=0$. Having found $C_1$, we have determined, up to $A_s$, the boundary layer solution $\pi (x)$. Now we match this solution with the fast-mode WKB solution $a_f(n) e^{-S(n)}$ in their joint region of validity $l_{2}\ll n_{2}-n \ll n_{2}$. To this end we expand $S(n)$ from Eq.~(\ref{n170}) around $n=n_2$ up to $(n-n_2)^2$ and evaluate $a_f(n)$, given by Eq.~(\ref{n190}), at $n=n_2$. The matching yields
\begin{equation}\label{A_f}
    A_f=A_s(2\pi/\delta)^{1/2} \Omega (1+\delta) e^{S(n_2)}
\end{equation}
and determines, up to $A_s$, the complete WKB solution at $1\ll n<n_2$.

The WKB approximation breaks down at $n={\cal O}(1)$. To find the QSD in this region we return to Eq.~(\ref{n20}) and notice that, at $n \ll \gamma\Omega$,  $\pi_n$ grows rapidly with $n$, so that $\pi_{n-1} \ll \pi_n$. The leading-order terms here are the following: $(n+1) \pi_{n+1}-(\gamma\Omega/2)\pi_n\simeq0$. That is, immigration and death balance each other and dominate over the reproduction.
The resulting recursion relation yields a Poisson distribution:
\begin{equation}
\pi_n=  \frac{\pi_0}{n!}\, \left(\frac{\gamma\Omega}{2}\right)^n\,.
\label{fr100}
\end{equation}
To determine the unknown constant $\pi_0$ we can match the asymptote (\ref{fr100}) with the asymptote
of the WKB solution $a_f(n) e^{-S(n)}$ at $1\ll n \ll \gamma\Omega$. To this end we expand the action $S(n)$ at $n \ll \gamma\Omega$: $S(n) \simeq  -n+n \ln\left[2n/(\gamma\Omega)\right]$.
On the other hand, at $n\gg 1$ we can use Stirling's formula $n!\simeq \sqrt{2 \pi n} \,(n/e)^n$ in Eq.~(\ref{fr100}). The matching yields
\begin{equation}
\pi_0 = \frac{2 \pi A_s \, e^{S(n_2)}}{\Omega \delta^{1/2} (1-\delta)} \,.
\label{fr105a}
\end{equation}
By now we have found, up to the normalization constant $A_s$, the QSD for \textit{all} $n$. The normalization, in the leading order, is determined by the region of $|n-n_{1}|\ll n_{1}$, where the fast-mode solution $\pi_n=a_f(n) e^{-S(n)}$ is approximately gaussian:
\begin{equation}\label{gaussian}
    \pi_n\simeq \frac{\sqrt{\pi} A_s\,(1+\delta)\, e^{\Omega\Delta s}}
    {\sqrt{2} \,\Omega^{3/2}\delta^{1/2}(1-\delta)^{3/2}} \,\exp\left[\frac{-(n-n_1)^2}{l_{1}^2}\right]\,.
\end{equation}
Here we have denoted $l_1=\left[2 \Omega (1/\delta-1)\right]^{1/2}$ and
\begin{eqnarray}
&&\hspace{-6mm}\Delta s (\delta) = [S(n_2)-S(n_1)]/\Omega = 2\delta \nonumber \\
&&\hspace{-6mm}-2\sqrt{1-\delta^2}\left(\arctan\sqrt{\frac{1+\delta}{1-\delta}}-
\arctan\sqrt{\frac{1-\delta}{1+\delta}}\right).
  \label{DeltaS}
\end{eqnarray}
Normalizing the gaussian~(\ref{gaussian}) to unity, we obtain
\begin{equation}\label{As}
    A_s=\frac{\Omega \delta (1-\delta)}{\pi (1+\delta)}\,e^{-\Omega\Delta s}
\end{equation}
which completes our calculation of $\pi_n$ for all $n$. Figure 1 shows the resulting QSD
for $\Omega=100$ and $\gamma=1/2$.
\begin{figure}[ht]
\includegraphics[width=2.5in,clip=]{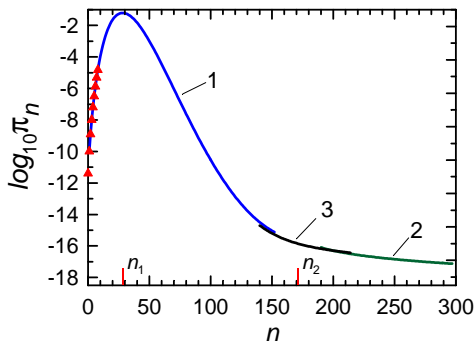}
\caption{(color online). The quasi-stationary distribution (QSD) for $\Omega=100$ and $\gamma=1/2$.
The QSD includes four overlapping asymptotes: the fast-mode (1), the slow-mode (2), the boundary-layer (3), and the Poisson distribution~(\protect\ref{fr100}) (triangles).
The small mismatch between the curves is due to higher-order corrections. We found numerically that the mismatch goes down as $\sim \Omega^{-1/2}$ at large $\Omega$.} \label{fig20}
\end{figure}

\textit{Decay time.} Having found the QSD we can calculate the decay time $\tau$. We return to Eq.~(\ref{n20}) and sum it up over $n$ from zero to infinity. By virtue of Eq.~(\ref{n150}), the first term on the right tends to $A_s/(2 \Omega)$, while the rest of the terms tend to zero. We obtain
\begin{equation}
E\sum\limits_{n=0}^\infty \pi_n=E=\frac{A_s}{2\Omega} = \frac{\delta (1-\delta)}{2\pi (1+\delta)}\,e^{-\Omega \Delta s}\,,
\label{m80}
\end{equation}
which is exponentially small as long as $\Omega \Delta s \gg 1$. The decay time, in physical units, is therefore
\begin{equation}
\tau=\frac{1}{\mu E}=\frac{2\pi(1+\delta)}{\mu\delta\, (1-\delta)}\,
e^{\Omega \Delta s}\,.
\label{fr10}
\end{equation}
$\Delta s$ is monotone increasing with $\delta$; its asymptotes are
$$
\Delta s=\left\{\begin{array}{ll}
(2/3)\delta^3 + (4/15) \delta^5+\dots,\;\;\;\delta\ll 1\,, \\
 2-\pi\sqrt{2 (1-\delta)}+\dots,\;\;\;1-\delta\ll 1\,.
\end{array}
\right.
$$
The exponent $e^{\Omega \Delta s}$ in Eq.~(\ref{fr10}) coincides with that obtained by Elgart and Kamenev \cite{Kamenev1} who only considered the leading order of
(a different version of) WKB. Our result (\ref{fr10}) goes beyond the leading order and includes a pre-exponent. The pre-exponent diverges as $\delta\to 1$ (when the immigration is stopped), so the decay time $\tau$ diverges. This result could not have been predicted in the leading order of WKB \cite{Kamenev1}.

The assumptions made in the process of derivation of our results include the strong inequalities $\Omega \Delta s>>1\,,$ $\,l_{2} << n_{2} - n_{1}$ and $l_{1} << \mbox{min}\, (n_{1}, \,n_{2} - n_{1})$.  For $\delta ={\cal O}(1)$, all the assumptions holds for $\Omega \gg 1$. For $0<\delta\ll 1$ [above but close to the bifurcation point of the birth of the fixed points $n_1$ and $n_2$ of Eq.~(\ref{rateeq})] the criterion is more stringent: $\Omega \delta^3 \gg 1$.

\textit{In conclusion}, by using a simple birth-death process as an example, we have developed a systematic theory of unlimited population growth driven by demographic noise. We have found the complete metastable probability distribution of the population at long times, $P_n(t) \simeq \pi_n \, e^{-t/\tau}$, and the previously unknown important pre-exponential factor in the decay time $\tau\sim e^{-\Omega \Delta s}$.  As $\pi_n$ has a power-law tail $\sim n^{-2}$, no distribution moments, except the zeroth one, exist. As a result, the initial-value problem for the master equation (\ref{p10}) is highly singular. When starting from a well-behaved $P_n(t=0)$, all of the distribution moments, except the zeroth one, diverge already at $t>0$. 

A general outcome of this work is that slow WKB modes should play an important role, along with fast WKB modes, in population escape problems. Consider, as an example, the Schl\"{o}gl model \cite{schloegl} where, in addition to our three reactions, one also has $3A\stackrel{\nu}{\rightarrow} 2A$. For very small $\nu$ the rate equation has an additional attracting point $n_3 \gg n_2$.  Here the stochastic population switches randomly between two metastable states peaked at $n_1$ and $n_3$. Now, if $P_n(t=0)$ is located around $n=n_1$, there is an exponentially long intermediate regime when the probability flux is directed from $n=n_1$ to $n=n_3$, whereas the reverse flux is negligible. This important regime is accurately captured by the solution presented above. The slow-mode component of the solution (which describes deterministic motion ``down the hill") is vital in determining the pre-exponents of the QSD and of the decay time.

We acknowledge a useful discussion with Alex Kamenev.  This work was supported by the Israel Science
Foundation (Grant No. 408/08).

\end{document}